\documentclass[usenatbib]{mn2e}
\usepackage{epsfig}
\usepackage{graphicx}
\usepackage{amssymb}
\usepackage{aas_macros}
\usepackage{color}

\newcommand{\oversim}[2]{\protect{\mbox{\lower0.5ex\vbox{%
   \baselineskip=0pt\lineskip=0.2ex
   \ialign{$\mathsurround=0pt #1\hfil##\hfil$\crcr#2\crcr\sim\crcr}}}}} 
\newcommand{\simgreat}{\mbox{$\,\mathrel{\mathpalette\oversim>}\,$}} 
\newcommand{\simless} {\mbox{$\,\mathrel{\mathpalette\oversim<}\,$}} 

\begin{document}

\title[Impact of cusps and cores on the satellite galaxy population]{The impact of dark matter cusps and cores on the satellite galaxy population around spiral galaxies}

\author[Pe\~{n}arrubia et al.]{Jorge Pe\~{n}arrubia$^{1,2}$\thanks{jorpega@ast.cam.ac.uk}, Andrew J. Benson$^{3}$, Matthew G. Walker$^{1}$, Gerard Gilmore$^{1}$,  
\newauthor  
Alan W. McConnachie$^{4}$ \& Lucio Mayer$^{5}$  \\
$^{1}$Institute of Astronomy, University of Cambridge, Madingley Road, Cambridge, CB3 0HA, UK\\
$^{2}$Kavli Institute for Cosmology, Madingley Road, Cambridge CB3 0HA, UK\\
$^{3}$California Institute of Technology, Pasadena, CA 91125, USA\\
$^{4}$NRC Herzberg Institute of Astrophysics, 5071 West Saanich Road, Victoria V9E 2E7, Canada \\
$^{5}$Institute for Theoretical Physics, University of Zuerich, Winterthurerstr. 190, 8057, Zurich, Switzerland\\
}

 \maketitle  

\begin{abstract} 
We use N-body simulations to study the effects that a divergent (i.e. ``cuspy'') dark matter profile introduces on the tidal evolution of dwarf spheroidal galaxies (dSphs). Our models assume cosmologically-motivated initial conditions where dSphs are dark-matter dominated systems on eccentric orbits about a host galaxy composed of a dark halo and a baryonic disc. We find that the resilience of dSphs to tidal stripping is extremely sensitive to the cuspiness of the inner halo profile; whereas dwarfs with a cored profile can be easily destroyed by the disc component, those with cusps always retain a bound remnant, even after losing more than 99.99\% of the original mass. For a given halo profile the evolution of the structural parameters as driven by tides is controlled solely by the total amount of mass lost. 
This information is used to construct a semi-analytic code that follows the tidal evolution of individual satellites as they fall into a more massive host, which allows us to simulate the hierarchical build-up of spiral galaxies assuming different halo profiles and disc masses. 
We find that tidal encounters with discs tend to decrease the average mass of satellite galaxies at all galactocentric radii. 
Of all satellites, those accreted before re-ionization ($z\simgreat 6$), which may be singled out by anomalous metallicity patterns, provide the strongest constraints on the inner profile of dark haloes. These galaxies move on orbits that penetrate the disc repeatedly and survive to the present day {\it only} if haloes have an inner density cusp. 
We show that the size-mass relationship established from Milky Way (MW) dwarfs strongly supports the presence of cusps in the majority of these systems, as cored models systematically underestimate the masses of the known Ultra-Faint dSphs. 
Our models also indicate that a massive M31 disc may explain why many of its dSphs with suitable kinematic data fall below the size-mass relationship derived from MW dSphs. We also examine whether our modelling can constrain the mass threshold below which star formation is suppressed in dark matter haloes. We find that luminous satellites must be accreted with masses above $10^8$--$10^9M_\odot$ in order to explain the size-mass relation observed in MW dwarfs. 
\end{abstract} 

%

\section{Introduction}
One of the strongest predictions from our present cosmological paradigm, Cold Dark Matter (CDM), refers to the universal density profile shared by dark matter haloes on all mass scales (Navarro, Frenk \& White 1996, 1997; hereinafter NFW). These authors suggested a simple formula to describe the spherically averaged profile of dark matter haloes
\begin{equation}
\rho(r)=\frac{\rho_0}{(r/R_s)[1 + (r/R_s)]^2}. 
\label{eq:rhonfw}
\end{equation}

This ``cuspy'' profile diverges towards the centre of the halo as $\rho(r)\propto r^{-1}$, although the enclosed mass is finite. Subsequent work has confirmed the absence of central-density cores in dark matter simulations (Moore et al. 1999b, Ghigna et al. 2000; Fukushige \& Makino 2001; Gao et al. 2004; Navarro 2004; Diemand et al. 2005; Springel et al. 2008a). 

Despite this consensus, there have been conflicting claims about the slope of the inner halo profile, $\gamma\equiv-{\rm d}\ln \rho/{\rm d}r$. For example, Diemand et al. (2004, 2005) claims a central slope of  $\gamma=1.2$, whereas other authors argued in favour of steeper, $\gamma=1.5$, cusps (Moore et al. 1999b; Gighna et al. 2000). Recently, Navarro et al. (2010) have found small, but significant deviations from an NFW profile in the inner-most regions of different Milky Way-size halo realizations (the Aquarius project; Springel et al. 2008b). This result may suggest that the slope of the inner halo profile may not strictly be ``universal'' (see also Gao et al. 2008), which might potentially explain the discrepancy found in the literature about the value of $\gamma$. 

The existence of cores or cusps in the centre of CDM haloes has important observational consequences for the discovery and interpretation of possible signals of dark matter annihilation in gamma-ray surveys (Stoehr et al. 2003; Diemand et al. 2007; Kuhlen et al. 2008; Springel et al. 2008b), since the strength of the signal goes with the square of the dark matter density. 

Interestingly, the presence of cusps in dark matter haloes has been challenged by measurements of the rotation curves of Low Surface Brightness galaxies (LSBs). Observations of LSBs, which are supposed to be embedded in a dark matter-dominated potential, seem to favour constant-density cores, $\gamma=0$, for most observed systems (e.g. Flores \& Primack 1994; Salucci \& Burkert 2000; Swaters et al. 2003; Simon et al. 2003; Trachternach et al. 2008; Oh et al. 2008; de Block 2008, 2010). 

  Dwarf Spheroidal galaxies (dSphs) are another example of dark matter-dominated galaxies where significant constraints on $\gamma$ can potentially be derived from the kinematics of their stellar constituents. Two types of analysis can be found in the literature: the first solves the Jeans' equations to  estimate the mass enclosed within the stellar radius. Unfortunately, in pressure supported systems  mass and orbital anisotropy are degenerate, making any deductions of the inner mass profile being cored or cuspy in general model-dependent (Koch et al. 2007a,b; Gilmore et al. 2007, Evans et al. 2009; Walker et al. 2009). A second analysis which aims at breaking the above degeneracy appeals to full multi-component distribution-function models, which fit the intrinsic and projected moments from hundreds of discrete radial velocities. Investigations using this technique seem to favour the presence of cores in the dark matter haloes of dSphs (Wilkinson et al. 2002; Gilmore et al. 2007). 

The apparent difficulties of CDM to reproduce observations on small scales using pure-dark matter simulations has not deterred a rich body of work from postulating alternative explanations within the present paradigm. Several authors have argued that baryons may play a fundamental role in sculpting the inner profiles of dark matter haloes, even in galaxies that at present are heavily dark matter-dominated. For example, mass outflows driven by supernova explosions may under plausible assumptions remove dark matter cusps (e.g. Navarro, Vincent \& Frenk 1996; Read \& Gilmore 2005; Governato et al. 2010). Also, massive clusters that sink into the inner-most regions of dark matter haloes via dynamical friction will exchange momentum with central particles, ejecting them from the cusp and creating a core in the process (Goerdt et al. 2008).

It is however debatable whether these processes are relevant on dSph scales, given that only a couple of MW dwarfs (Sagittarius and Fornax) hold a bound cluster population, and that the large majority of dwarfs with suitable kinematics show mass-to-light ratios $\simgreat 100$. 

Clues to these questions may be gathered from examining the overall properties of the satellite galaxy {\it population}, rather than from individual systems. 
This paper is aimed at studying the influence of dark matter cusps and cores on the evolution of satellite galaxies driven by tides, as well as exploring whether the satellite population around spirals may retain information on the inner structure of dark matter haloes. The arrangement of the paper is as follows: \S\ref{sec:models} presents the models assumed to describe the host and satellite galaxies, as well as the numerical set-up. \S\ref{sec:tidev} presents the results of our N-body experiments regarding the influence of cusps and cores on the dynamical evolution of dSphs. Appendix\ref{sec:sa_code} uses these results to construct a semi-analytic algorithm that simulates the formation of spiral galaxies through the accretion of individual subhaloes. \S\ref{sec:sa} shows the results of applying this code to Milky Way-like galaxies and analyzes the properties of the surviving satellite population for a range of inner halo slopes and disc masses. We end with a brief summary in \S\ref{sec:summary}.

\section{Galaxy models}\label{sec:models}

\subsection{Density profiles}\label{sec:prof}
We assume that dark matter haloes follow a universal density profile that can be fitted by the general five-parameter formula (Zhao 1996; Kravtsov et al. 1998)
\begin{eqnarray}
\rho(r)=\frac{\rho_0}{(r/R_s)^\gamma[1 + (r/R_s)^\alpha]^{(\beta-\gamma)/\alpha} } ~~~(r\leq R_{\rm vir}); 
\label{eq:rho}
\end{eqnarray}
where $\gamma$ controls the inner slope of the profile, $\beta$ the outer profile, and $\alpha$ the transition between both. In cosmological simulations, the characteristic density $\rho_0$ and scale radius $R_s$ are sensitive to the epoch of formation and correlate with the halo virial radius via the concentration parameter $C_{\rm vir}\equiv R_{\rm vir}/R_s$. 

The virial radius is defined so that the mean over-density relative to 
the critical density is $\Delta$, 
\begin{equation} 
{M_{\rm vir} \over (4/3) \pi R_{\rm vir}^3}= \Delta \ \rho_{\rm crit}, 
\end{equation} 
where $\rho_{\rm crit}=3 H_0^2/8 \pi G$, and $H_0=100\, h$ km/s/Mpc is 
the present day value of Hubble's constant. 
 
The choice of $\Delta$ varies in the literature, with some authors 
using a fixed value, such as  NFW, who adopted $\Delta=200$, and others 
who choose a value motivated by the spherical collapse model, where 
(for a flat universe) $\Delta\sim 178\, \Omega_{\rm m}^{0.45}$ (Lahav 
et al 1991, Eke et al 1996). The latter gives $\Delta=95.4$ at $z=0$ 
in the concordance $\Lambda$CDM cosmogony, which adopts the following 
cosmological parameters: $\Omega_{\rm m}=0.3$, $\Omega_\Lambda=0.7$, 
$h=0.7$, consistent with constraints from CMB measurements and galaxy 
clustering (see, e.g., Spergel et al 2007 and references therein). 

The density profile given by eq.~[\ref{eq:rho}] may diverge as $r \rightarrow 0$ if the inner slope $\gamma > 0$. In the literature such dark matter haloes are called ``cuspy'', and a prototypical example is the NFW profile, which corresponds to $(\alpha, \beta, \gamma)=(1,3,1)$. In contrast, ``cored'' haloes have a converging density profile at small radii, i.e. $\gamma=0$.

Also, density profiles with outer slopes $\beta> 3$ lead to finite mass models, but for $\beta\leq 3$ the cumulative mass profiles diverges as $r \rightarrow \infty$. To avoid that problem, Kazantzidis et al. (2004) impose a truncation at $R_{\rm vir}$ that can be written as

\begin{eqnarray}
\rho(r)=\frac{\rho_0}{C_{\rm vir}^\gamma (1+C_{\rm vir}^\alpha)^{(\beta-\gamma)/\alpha}}\bigg(\frac{r}{R_{\rm vir}}\bigg)^\epsilon \exp\bigg(-\frac{r-R_{\rm vir}}{R_{\rm dec}}\bigg)~~~(r>R_{\rm vir});
\label{eq:rhotrun}
\end{eqnarray}
where $R_{\rm dec}$ is a small quantity, whose value we choose as $0.1 R_{\rm vir}$. 

To obtain a continuous logarithmic slope $\epsilon$ is defined as

\begin{eqnarray}
\epsilon = \frac{-\gamma -\beta C_{\rm vir}^\alpha}{1+C_{\rm vir}^\alpha} + \frac{R_{\rm vir}}{R_{\rm dec}}.
\label{eq:eps}
\end{eqnarray}

Our models are purely gravitational and therefore scale-free. When 
convenient for interpretive purposes, we shall scale our numerical 
units to physical parameters assuming that the host galaxy potential 
has $M_{\rm vir}=10^{12} M_\odot$, $R_{\rm vir}=258$ 
kpc and $R_s=21.5$ kpc, assuming $\Delta=101$. These parameters are thus similar to those estimated by Klypin et al. (2003) for the MW dark matter halo.

To explore the effects that the presence of a baryonic disc in the host galaxy induces on the satellite galaxy population, we add a new force component ${\bf f}_d=-\nabla \Phi_d$ in the form of a Miyamoto-Nagai (1975) disc model
\begin{equation}
\Phi_d(R,z)=-\frac{G M_d }{\sqrt{R^2 + (a+\sqrt{z^2+b^2})^2}};
\label{eq:phid}
\end{equation}
where $M_d$ is the disc mass, $a=6.5$ kpc and $b=0.25$ kpc are the radial and vertical scale lengths. 

\begin{figure}
  \includegraphics[width=84mm]{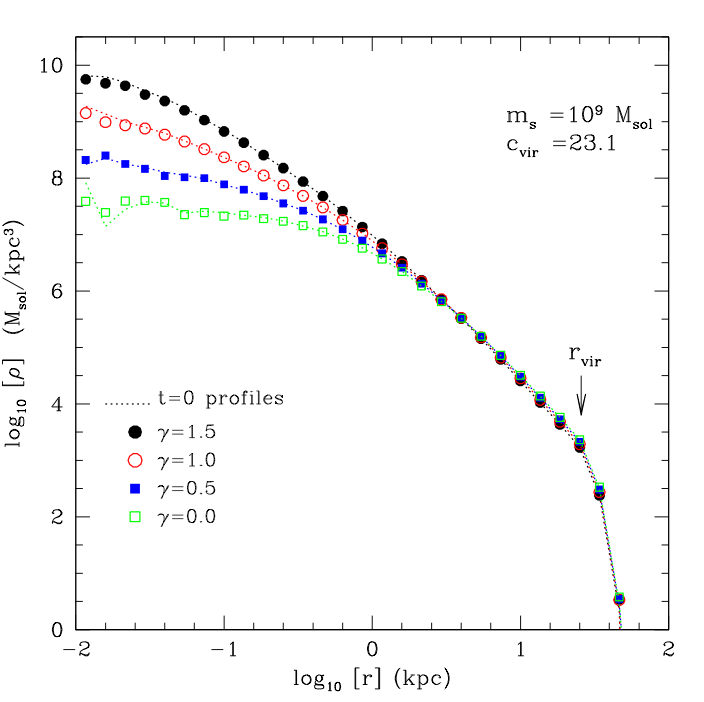}
\caption{Density profiles of dark matter halo models for different values of $\gamma$, and same virial mass and concentration. {\it Cuspy} and {\it cored} dark matter haloes are those with $\gamma > 0$ and  $\gamma = 0$, respectively. Dotted lines correspond to initial N-body realizations. Symbols show the same realizations integrated for 14 Gyr in isolation. The good agreement between lines and dots implies that our numerical setup is adequate and free from numerical artifacts induced by numerical limitations down to scales comparable with the grid-size of the highest-resolution zone used by {\sc superbox}, $r_s/126\simeq 8.8\times 10^{-3} $ kpc.}
\label{fig:prof}
\end{figure}

\subsection{N-body realizations of satellite galaxies}\label{sec:satmod}
Our satellite galaxy models are constructed by generating 5-million particle realizations of the profile outlined in eq.~[\ref{eq:rho}], using the code kindly made available by S. Kazantzidis (Kazantzidis 2004, 2006). These models are composed purely of dark matter, thus neglecting the contribution from baryons to the overall potential of satellite galaxies. This approximation is valid for the large majority of satellite galaxies, i.e. dwarf galaxies, which typically show very high mass-to-light ratios (Aaronson 1983, Aaronson \& Olszewski 1987; Mateo 1998; Walker et al. 2007; Simon \& Geha 2007; Strigari et al. 2008; Pe\~narrubia et al. 2008a).

We have chosen parameters for the dwarf consistent with those expected 
for dwarf galaxy halos (see, e.g., Pe\~narrubia et al. 2008a). Using the same physical scaling 
adopted for the host galaxy (\S~\ref{sec:prof}) we consider halo models with $m_{\rm 
vir}=10^9 M_\odot$, $r_{\rm vir}=25.7$ kpc, and $c_{\rm vir}=23.1$. In these models the scale radius is independent of the inner slope of the profile, $r_s=1.11$ kpc. 
We emphasize, however, that our models are scale free, so that the 
only relevant parameters in the calculations are the {\it relative} 
masses and sizes of the host and dwarf. In what follows, capital and small letters  
will be used to denote quantities corresponding to the ``host'' or 
the ``satellite'' models, respectively.

Fig.~\ref{fig:prof} shows the resulting density profiles of the satellite for four different values of the slope of the inner density profile, $\gamma$. Note that differences arise mostly within $r\leq r_s$. Although in those regions the density of dark matter can differ by several orders of magnitude depending on $\gamma$, in fact the differences in enclosed mass are considerably smaller, $m(<r_s)/m_{\rm vir}=3.86\times 10^{-2}, 5.74\times 10^{-2}, 8.69\times 10^{-2}$ and $13.45\times 10^{-2}$ for $\gamma=0.0, 0.5, 1.0$ and 1.5, respectively. In the following Sections we will see how these relatively small differences in central mass play a capital role in the overall evolution of satellite galaxies orbiting around spirals.

\subsection{The N-body code}
 
We follow the evolution of the dSph N-body model in the host potential 
using {\sc Superbox}, a highly efficient particle-mesh gravity code 
(see Fellhauer et al. 2000 for details). {\sc Superbox} uses a 
combination of different spatial grids in order to enhance the 
numerical resolution of the calculation in the regions of interest. 
In our case, {\sc Superbox} uses three nested grid zones centered on 
the highest-density particle cell of the dwarf.  This center is 
updated at every time step, so that all grids follow the satellite 
galaxy along its orbit. 
 
Each grid has $128^3$ cubic cells: (i) the inner grid has a 
spacing of $dx=r_s / 126\simeq 8\times 10^{-3} \, r_s$ and is meant to 
resolve the innermost region of the dwarf, where the effects of varying $\gamma$ are strongest.
(ii) The middle grid extends to cover the whole 
dwarf, with spacing $r_{\rm vir}/ 126$. (iii) The outermost grid 
extends out to $50\times r_{\rm vir}$ and is meant to follow particles 
that are stripped from the dwarf and that orbit within the main 
galaxy.
 
{\sc Superbox} uses a leap-frog scheme with a constant time-step to 
integrate the equations of motion for each particle. We select the 
time-step according to the criterion of Power et al (2003); applied to 
our dwarf galaxy models, this yields $\Delta t=4.6$ Myr for the 
physical unit scaling adopted in Sec.~\ref{sec:prof}.

\subsection{Tests of the dSph N-body model} 
\label{ssec:tests} 
 
We have checked explicitly that our procedure for building the dSph 
model leads to a system that is in equilibrium and that does not 
evolve in isolation away from the prescribed configuration. This is shown 
in Fig.~\ref{fig:prof}, where we plot the initial density profiles of four dark matter models with different inner slopes (symbols).  Dotted lines show the same 
profiles, but after evolving the system in isolation for 14 Gyr with 
{\sc Superbox}. The good agreement between lines and symbols shows 
that our numerical choices are appropriate, and that the evolution of 
the N-body model, on the scales of interest, is free from artifacts
induced by the finite spatial and time resolution of the calculation.

\section{Tidal evolution of dark matter subhaloes}\label{sec:tidev}

\begin{figure}
  \includegraphics[width=84mm]{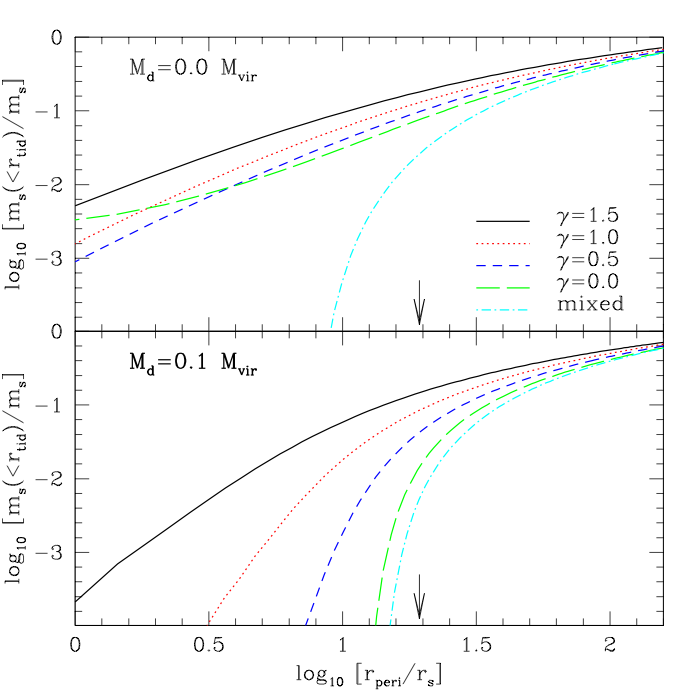}
\caption{Mass enclosed within the tidal radius (see text) of the satellite models presented in Fig.~\ref{fig:prof} estimated at different Galactocentric distances for host galaxies without ({\it upper panel}) and with a baryonic disc component with a fiducial mass $M_d=0.1 M_{\rm vir}$ ({\it lower panel}), respectively. The ``mixed'' models refer to satellites with cored $\gamma=0.0$ profiles orbiting about a host galaxy with a $\gamma=1.0$ halo profile. All radii are normalized by the scale radius $r_s$ of the satellite model. Arrows mark the position of the host scale radius, i.e. $R_s/r_s$, for this particular satellite model.}
\label{fig:rtrp}
\end{figure}

\subsection{Tidal strength}\label{sec:tidstr}
Satellite galaxies may lose some fraction of their initial mass to tides if their orbits bring them close to the host galaxy centre. We may quantify the expected strength of the tides by computing the ``tidal radius'' ($r_{\rm tid}$) of the satellite galaxy if this was placed on a circular orbit at pericentre

\begin{equation}
\langle \rho_{\rm sat}(r_{\rm tid}) \rangle =3 \langle \rho_{\rm host}(r_{\rm peri}) \rangle.
\label{eq:rtid}
\end{equation}

Here $\langle  \rho(r)\rangle$ denotes mean enclosed density within a radius $r$. Pe\~narrubia et al. (2008b) have shown that the total amount of mass, $m_{\rm tid} \equiv m_s(<r_{\rm tid})$, provides an accurate estimate of the amount of mass that the satellite may expect to retain after completing several orbits in the host potential. 

In order to explore the resilience to tides of galaxies with different $\gamma$-values we have calculated the mass enclosed within the tidal radius of the models shown in Fig.~\ref{fig:prof} as a function of pericentric distance using eq.~[\ref{eq:rho}] and~[\ref{eq:rtid}] and imposing a universal density profile (but different scaling) for {\it both} host and satellite models. The results are shown in Fig.~\ref{fig:rtrp}. In absence of a baryonic disc embedded in the host halo (i.e. $M_d=0$), cuspy and cored satellite models show similar resilience to tides, although cored satellites are slightly more prone to tidal stripping than cuspy ones. 
Satellites with divergent profiles show a monotonic decline of $m_{\rm tid}$ as the pericentric radius decreases. Interestingly, this is not the case for satellites with a finite central density, which can retain a small bound mass fraction $m_{\rm tid}\sim$const even if the orbits are fairly eccentric $r_{\rm peri}\simless r_s$. 

Motivated by recent claims that star formation/feedback processes in the inner regions of dark matter haloes can turn cusps into cores (e.g. Governato et al. 2010, and references therein) we have also explored a ``mixed'' case wherein our satellite models are cored and orbit a host halo with a $\gamma=1.0$ profile (dotted-dashed lines). Fig.~\ref{fig:rtrp} shows that in this configuration satellite galaxies would undergone tidal disruption if their orbits bring them at galactocentric distances $\simless 10 r_s$.

In models where the parent galaxy hosts a baryonic disc with a fiducial mass $M_d=0.1M_{\rm vir}$ ({\it lower panel}) the influence of the disc component increases with decreasing $\gamma$ at all pericentric radii.
This is because at small orbital pericentres, $r_{\rm peri}\simless R_s$, the mean density of spiral galaxies embedded in {\it cored} haloes scales as 
$$\langle \rho_{\rm host}\rangle \sim \rho_0 + 3a  M_d  /[8\pi (a+b)^3] r_{\rm peri}^{-1}.$$ 
Thus, in models with $\gamma=0$, the disc component dominates the host tidal field if $r_{\rm peri}\simless (3a [8\pi (a+b)^3]^{-1} M_d /\rho_0)$. Complete tidal disruption is assured if the mean density of the host is higher than the central density of the satellite. This occurs at $r_{\rm peri}\simless 9a M_d/[2(a+b)^3 m_s] r_s^3 y(c_{\rm vir})$, where $y(x)=\ln(1+x)-x(2+x)/[2(1+x)^2]$. Using the model parameters adopted in \S\ref{sec:models}, disruption is expected for orbits with $r_{\rm peri}\simless 32 r_s$.

In contrast, at small radii the mean density of spiral galaxies with cuspy, $\gamma=1$, haloes scales as 
$$\langle \rho_{\rm host}\rangle \sim 3\rho_0 R_s^2 r_{\rm peri}^{-2} +3a M_d / [8\pi (a+b)^3] r_{\rm peri}^{-1}.$$
 Thus in these models the disc tidal field does {\it not} dominate at small radii. Furthermore, in the limit $r_{\rm peri}\rightarrow 0$ one can readily show that eq.~[\ref{eq:rtid}] has an analytical solution, which can be written as $r_t=[m_s R_s f(c_{\rm vir})/ (3 r_s M_{\rm vir} f(C_{\rm vir})]^{1/2}r_{\rm peri}$, where $f(x)=\ln(1+x)-x/(1+x)$. Since at small radii $m_s(<r_{\rm tid})\propto r_{\rm tid}$,  this implies that $m_s(<r_{\rm tid})=0$ if and only if $r_{\rm peri}=0$, which suggests that only radial orbits may lead to full tidal disruption of satellites orbiting in such a galaxy potential. A similar prediction was made by Goerdt et al. (2007) using different analytical arguments. As we shall see below, the apparent ``indestructibility'' of cuspy subhaloes is confirmed by our N-body simulations.

It is clear from inspecting this Figure that baryonic discs may represent an important factor in the survival of satellite galaxies around spiral galaxies, and that the disc influence strongly varies depending on the slope of the inner density profile.

\begin{figure*}
\includegraphics[width=148mm]{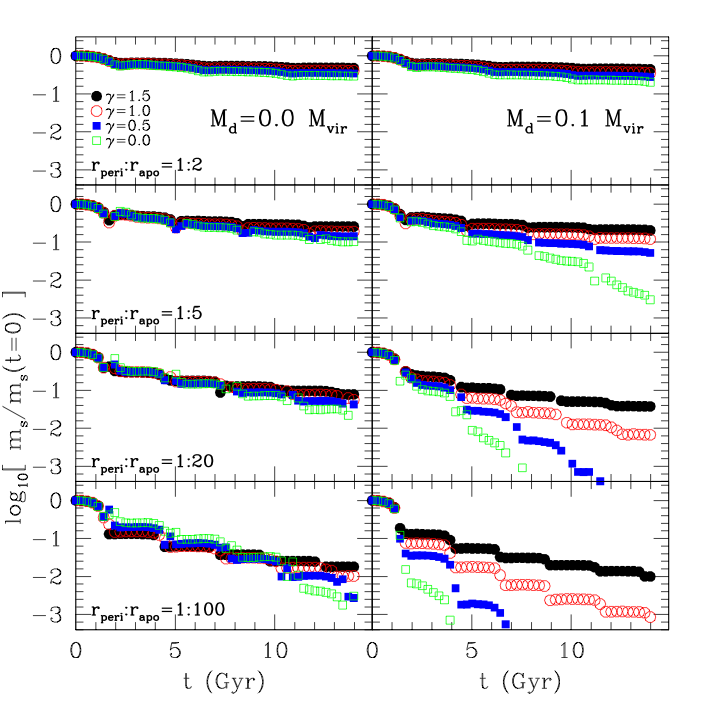}
\caption{Bound mass fraction as a function of time for different orbits and values of $\gamma$. Left panels show the mass evolution of subhaloes in a host halo without disc. Right panels show the effects of injecting a disc component in the host galaxy potential. Note that the disc influence on the survival of dark matter subhaloes strongly depends to the slope of the inner density profile, $\gamma$. In particular, the mass evolution of satellite galaxies with prominent cusps, i.e. $\gamma=1.5$, is barely altered by the disc presence, whereas cored satellites, i.e. $\gamma=0.0$, suffer severe mass loss if their orbits bring them into the disc vicinity. }
\label{fig:mass}
\end{figure*}

\subsection{The orbits}
To explore self-consistently the tidal evolution of satellite galaxies we place the N-body models outlined in \S\ref{sec:satmod} in orbit within the host galaxy assuming different combinations of apocentric/pericentric ratio. All orbits have apocentres $r_{\rm apo}\simeq 0.70 R_{\rm vir}=180$ kpc. We consider a large range of pericentric radii $r_{\rm peri}=1/2, 1/5, 1/20$ and $1/100$ of $r_{\rm apo}$. Note that the smallest pericentre is comparable to the scale radius of the satellite, and that this type of close-range encounter is rare in self-consistent cosmological simulations. In this contribution, this series of orbits can be regarded as one of increasing tidal strength for eccentric orbits of similar orbital period.
In all cases we follow the evolution of the satellite for $t_H=14$ Gyr in the scale units introduced in \S\ref{sec:prof}, which corresponds to 3--5 orbital periods.

\begin{figure*}
  \includegraphics[width=84mm]{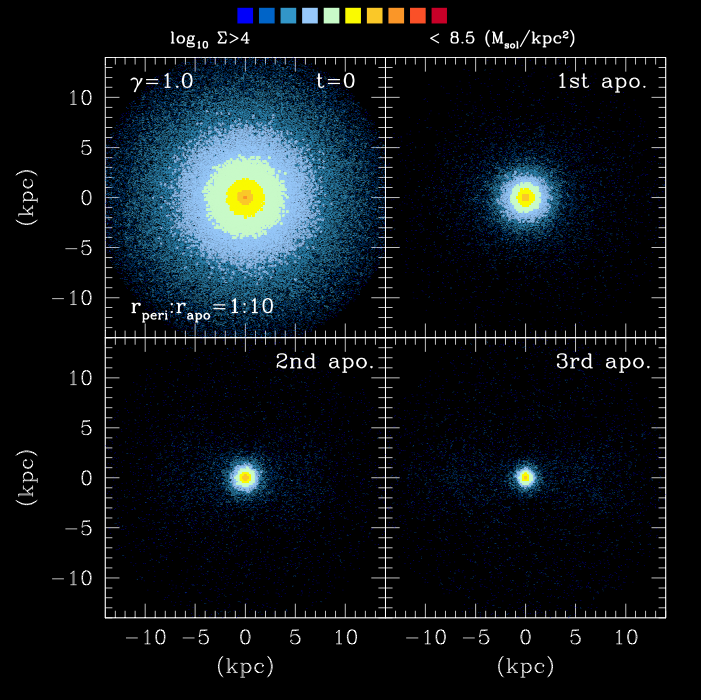}
\includegraphics[width=84mm]{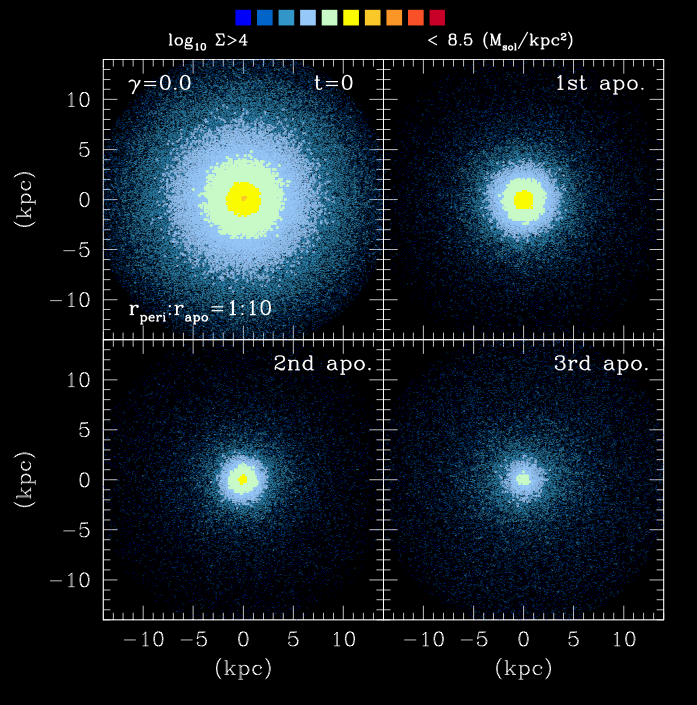}
\caption{Projection onto the orbital plane of a cuspy ($\gamma=1.0$, left panels) and cored ($\gamma=0.0$, right panels) satellite models that follow a highly eccentric orbit ($r_{\rm peri}:r_{\rm apo}=1:100$) about a host galaxy with no disc component. 
Each of the panels correspond to consecutive orbital apocentres for each model. After three consecutive pericentric passages, both satellites have lost 97\% of their initial mass. Particles have been colour-coded according to the value of the projected surface density at their location. Note how in both models the stripping proceeds from the outside in, and that the cuspy model shrinks at a faster pace than the cored one, without a substantial drop of the central density.}
\label{fig:xyz}
\end{figure*}

\subsection{Mass loss}
Tides are strongest at pericentre and trigger mass loss episodes that recur with every pericentric passage. The progressive stripping of satellites is shown in Fig.~\ref{fig:mass}, where dots represent the fraction of mass that remains bound to the satellite as a function of time. Pericentric passages are easily recognizable as the times when sudden drops in bound mass occur. Between pericentric passages the bound mass fraction remains nearly constant, suggesting, as indicated above, that tidal forces act in an impulsive regime and have little influence on the main body of the satellite at times other than pericentre.

 As expected from Fig.~\ref{fig:rtrp}, in the absence of a disc component in the host galaxy (left panel) the difference in mass loss that results from varying the slope of the inner density profile of our satellite models is in general small, and mostly arises at the latest stages of the orbital evolution. An exception can be gleaned from the most-eccentric modes plotted in Fig.~\ref{fig:mass} ($r_{\rm peri}:r_{\rm apo}=1:100$), which show that after the first pericentric interaction the bound mass fraction of satellites with prominent cusps ({\it filled circles}) is a factor 2--3 lower than that of the cored models ({\it open squares}). However, this situation reverses as the number of pericentric passages increases, which means that, when cored haloes begin to disrupt, they do so more rapidly than cuspy haloes. 

The right panel of Fig.~\ref{fig:mass} shows that the presence of a disc component in the host introduces remarkable differences between the mass evolution of cored and cuspy satellites. Whereas the rate at which satellites with prominent cusps, $\gamma=1.5$, lose mass is barely altered, the {\it full} tidal disruption of models with a $\gamma=0.0$ profile is assured if their orbits bring them into the vicinity of the disc. 

Indeed, dark matter haloes with $\gamma\geq 1.0$ show an extreme resilience to tides in all models explored here. As Fig.~\ref{fig:mass} shows, the first pericentric passage results in a significant decrease of the bound mass fraction. However, the mass loss rate progressively slows down as the number of consecutive pericentric passages increases, and the bound mass tends to converge to a finite value. This result implies that {\it dwarf galaxies embedded in haloes with strong cusps, $\gamma\geq 1.0$, cannot be fully disrupted by the tidal field of the host galaxy}, as expected from the analytical arguments given in \S\ref{sec:tidstr}\footnote{Note that Hayashi et al. (2003) stated that, if haloes are truncated at a radius $r< 0.77 r_s$, that would result in the total disruption of the remnant due to the positive binding energy of the NFW in that region. However, in \ref{sec:dens_evol} we show that tidal stripping does {\it not} truncate the density profile of dark matter haloes (see also Pe\~narrubia et al. 2008b, 2009)}. This has interesting consequences for the properties of the satellite galaxy population, an issue to which we return below.

The tidal mass stripping of dark matter haloes proceeds gradually from the outside in (e.g. Stoehr et al. 2002; Kazantzidis et al. 2004; Kampakoglou \& Benson 2007; Pe\~narrubia et al. 2008b). This is illustrated in Fig.~\ref{fig:xyz}, where we plot, at consecutive apocentres of the orbit, the position of the satellite particles in the satellite frame. This clearly shows how a halo is gradually stripped from the outside in, following an ``onion-peel'' process in which mass loss affects predominantly the outer, low-density regions of the satellite galaxy.

By colour-coding the particles according to the value of the projected surface density at their location we can appreciate fundamental differences in the evolution of dark matter haloes depending on whether they are cuspy (left panel) or cored (right panel), even if both satellite models have lost the same fraction of their original mass, 97\%, after three pericentric passages.
The first interesting difference is that the spatial sizes of the cuspy satellite models suffer a dramatic decrease as a result of mass loss. Yet the bound remnants show densities that have been barely altered by the action of tides. 
The response of cored satellite models to tidal mass loss is practically the opposite. In these systems the change in size if much less pronounced, and the density drops more strongly than in the cuspy models at all radii, an issue that we explore in more detail in the following Section.

\subsection{Evolution of the density profile}\label{sec:dens_evol}
\begin{figure}
  \includegraphics[width=84mm]{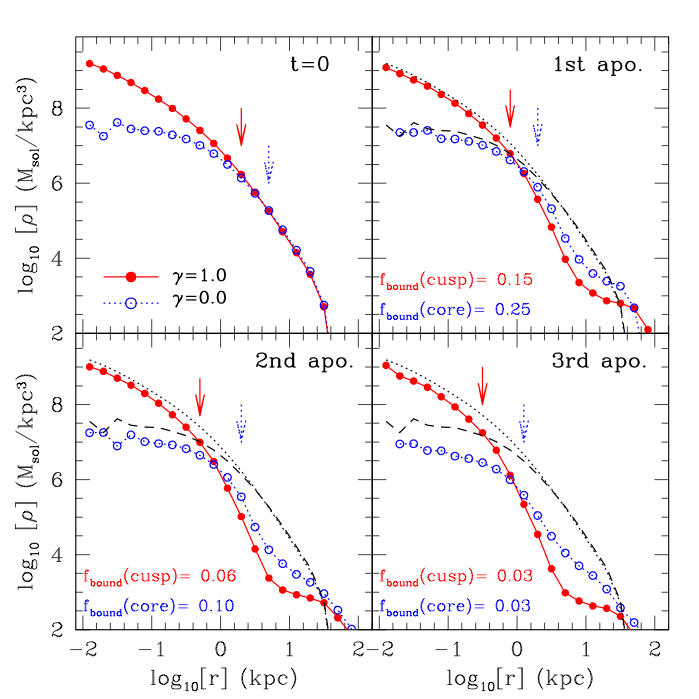}
\caption{ Halo density profile of the cuspy and cored halo models shown in Fig.~\ref{fig:xyz}. For comparison we also plot the initial density profiles in each panel, where dotted and dashed lines denote haloes with $\gamma=1.0$ and 0.0, respectively. The quantity $f_{\rm bound}$ refers to the bound mass fraction. Arrows indicate the position of $r_{\rm max}$, i.e. the radius where the circular velocity curve peaks. Note that the density profile in the inner regions of dark matter haloes is preserved even when large fractions of the initial mass are lost to tides.}
\label{fig:rhoevol}
\end{figure}

Fig.~\ref{fig:rhoevol} shows the evolution of the density profile of cored ($\gamma=0.0$, {\it open circles}) and cuspy ($\gamma=1.0$, {\it filled circles}) satellite models for the most extreme orbit ($r_{\rm peri}:r_{\rm apo}=1:100$) probed in our series. The systems are shown at three consecutive orbital apocentres, when the satellites have had some time to find dynamical equilibrium after being perturbed by the host tidal field at pericentre. The dotted/dashed lines are the same in each panel and show the initial density profiles for ease of reference. 

A few interesting points can be gleaned from this Figure. One is that it is possible to identify two distinct regions on the basis of the shape of the density profile: an inner region where the profile varies smoothly with radius, and an outer region where an ``excess'' of material is present with respect to a naive extrapolation of the inner profile. Several authors (e.g Oh, Lin \& Aarseth 1995; Mayer et al. 2002; Johnston et al. 2002) have shown that this type of ``breaks'' in the density profile are transient features that
arise from the presence of unbound particles in the outskirts of the system as they escape from the satellite. More recently, Pe\~narrubia et al. (2009) have shown that the position of the ``break'', $r_b$, marks the radius at which the local crossing time exceeds the time elapsed since pericentre, $t-t_p$, so that $r_b\propto \sigma (t-t_p)$, where $\sigma$ is the mean velocity dispersion of the satellite. 
Particles inside the break radius are thus within the region where enough time has elapsed for equilibrium to be reached. 

A second interesting point of note is that the evolution of the density profile is remarkably sensitive to the shape of the inner profile. Tides acting on cuspy haloes tend to induce strong drops of density in the outer regions, as expected from Fig.~\ref{fig:xyz}, whilst keeping the shape of the inner profile barely altered, even after the satellite has been stripped of 97\% of its initial mass. The bottom-right panel shows that, for a similar mass loss, tides tend to lower the central density of cored haloes more severely than in the case of haloes with a centrally-concentrated profile. 

The evolution of the satellite's circular velocity curve has a different amplitude depending on the slope of the inner halo profile. This is shown in Fig.~\ref{fig:rhoevol}, in which we mark the position where the circular velocity reaches its peak,$r_{\rm max}$, at consecutive orbital apocentres . As this Figures shows, mass loss results in a progressive shift of $r_{\rm max}$ towards the satellite centre, a process that is more pronounced in the case of cuspy satellite models. We inspect this issue in some detail in the following Section.

\subsection{Evolution of structural parameters}
As shown by Hayashi et al. (2003) and Pe\~narrubia et al. (2008b), the evolution of the structural parameters of NFW haloes as they are tidally stripped does solely depend on the {\it total amount of mass lost}, and not on the details of {\it how} the mass is stripped. 

\begin{figure}
  \includegraphics[width=84mm]{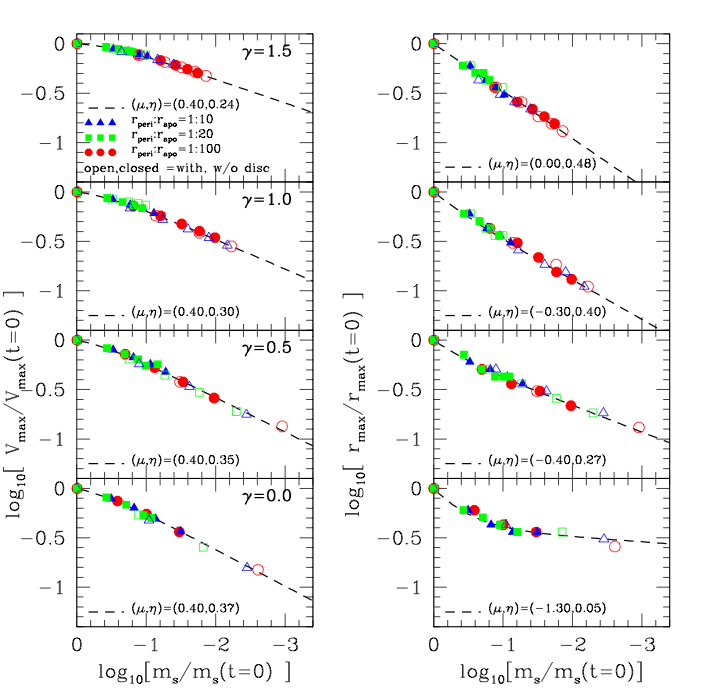}
\caption{ Evolution of the peak circular velocity $v_{\rm max}\equiv V_c(r_{\rm max})$ and its location, measured at the apocentres of the various orbits in our simulation series. Open and closed symbols denote simulations with and without a baryonic disc component ($M_d=0.1 M_{\rm vir}$), respectively. The dashed lines are our fits to the evolutionary tracks using eq.~[\ref{eq:g}]. }
\label{fig:vmrm}
\end{figure}

Fig.~\ref{fig:vmrm} shows that this result is generic to dark matter haloes that follow the density profile outlined by eq.~[\ref{eq:rho}]. The left and right panels of the Figure show respectively the fractional change of the peak velocity, $v_{\rm max}$, and its location, $r_{\rm max}$, as a function the satellite bound mass fraction $f_{\rm bound}\equiv m_s/m_s(t=0)$ for three of the four orbits surveyed in our study. Different symbols correspond to different orbits and are shown at consecutive orbital apocentres when the main body of the satellite galaxy is close to dynamical equilibrium.
Open and filled symbols denote models in which a disc component has been embedded in the parent galaxy. 

This Figure illustrates a few interesting differences in the evolution of satellite galaxies depending on how distended the halo profile is in the inner-most regions. The first is that, at a given mass loss fraction, the peak velocity declines more significantly as we decrease the value of $\gamma$. The second difference is in the evolution of the satellite size, which is clearly more substantial for satellite galaxies with cusps, as expected from Fig.~\ref{fig:xyz} and~\ref{fig:rhoevol}. The right panel of Fig.~\ref{fig:vmrm} shows that $r_{\rm max}$ is severely shortened by the action of tides if the slope of the inner halo profile is $\gamma>0$. Interestingly, the location of the peak velocity of cored models cannot drop further than a factor $\simeq 3$, regardless of the amount of mass lost to tides. This is in marked contrast with cuspy haloes, which may shrink into oblivion if tidal stripping carries on indefinitely.

Irrespective of pericentre, number of completed orbits, host potential, or slope of the inner density profile, the evolution of the structural parameters depends solely on how much the satellite mass has varied. This is a remarkable result, especially because our series of models includes extreme cases in which the satellite models lose more than 99.99\% of their original mass. Following Pe\~narrubia et al. (2008b), we use a simple empirical formula to fit the evolution of the structural parameters
\begin{equation}
g(x)=\frac{2^\mu x^\eta}{(1+x)^\mu};
\label{eq:g}
\end{equation}
where $x\equiv m_s/m_s(t=0)$ and $g(x)$ represents either $v_{\rm max}$ or $r_{\rm max}$. The best-fit values of $\mu$ and $\eta$ have been plotted in each of the panels of Fig.~\ref{fig:vmrm}, and the results are shown with dashed lines.

The fact that the evolution of satellite galaxies as they are stripped by tides follows well-defined {\it tracks} will be used in the following Sections to build a simple semi-analytical algorithm that follows the disruption of dark matter haloes with different inner profiles with reasonable accuracy (see Appendix A).

\section{Semi-analytic realizations of galactic merger trees}\label{sec:sa}

The results presented in \S\ref{sec:tidev} show that the response of galaxies to tides strongly depends on the slope of the inner density profile of dark matter haloes, $\gamma$, as well as to the presence of a baryonic disc embedded in the host galaxy.  This raises the question of whether the satellite populations surrounding spiral galaxies may provide insights into the inner structure of dark matter haloes. In order to explore this issue, simulations that follow the hierarchical formation of spiral galaxies through the accretion of individual satellites are required. We begin by outlining a method to construct merger trees of spiral galaxies, and then discuss the effects of varying $\gamma$ and the disc-to-parent halo mass ratio on the present population of satellite galaxies.

\subsection{Build-up of merger trees}\label{sec:mtree}

\subsubsection{The host galaxy}
\begin{figure}
  \includegraphics[width=84mm]{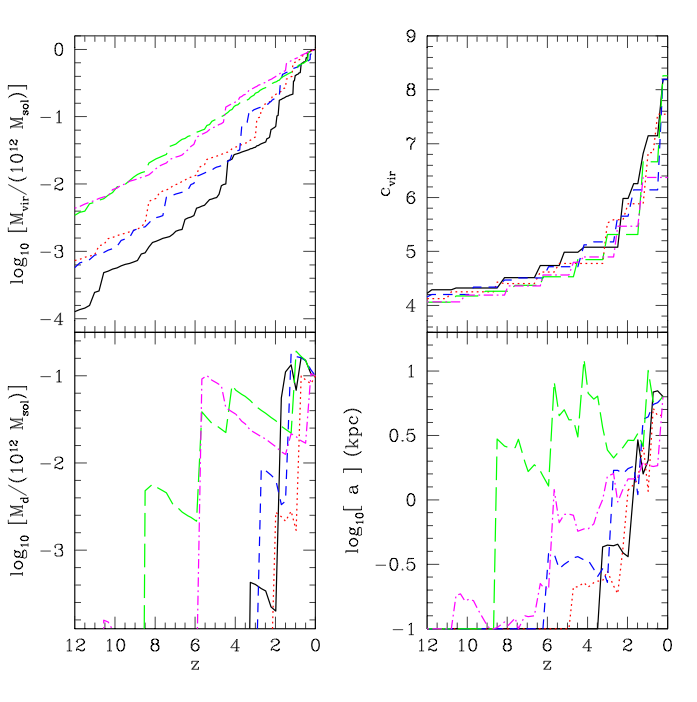}
\caption{ The evolution of the host galaxy parameters for five different merger tree realizations of a halo with $M_{\rm vir}=10^{12} M_{\odot}$ at $z=0$ is shown in the upper panels. Here we plot the virial mass and concentration in the upper-left and -right panels, respectively.  The disc component embedded in the host galaxy also grows with time. For this particular plot we show the evolution of a disc with $M_d=0.1 M_{\rm vir}$ and $a=6.5$ kpc at $z=0$. The disc mass and size are plotted in the lower-left and -right panels, respectively. }
\label{fig:hostevol}
\end{figure}

We construct merger trees using Monte Carlo methods. Specifically we employ 
the merger tree algorithm described by Parkinson et al. (2008) which 
is itself an empirical modification of that described by 
Cole et al. (2000). We adopt the parameters $(G_0,\gamma_1,
\gamma_2)=(0.57,0.38,-0.01)$ that are found 
to provide the best fit\footnote{Benson (2008) found 
an alternative set of parameters which provided a better match to the 
evolution of the overall halo mass function but performed slightly less well 
(although still quite well) for the progenitor halo mass functions. We have 
chosen to use the parameters of Parkinson et al. (2008) as 
we wish to get the progenitor masses as correct 
as possible.} to the statistics of halo progenitor masses measured from the 
Millennium Simulation by Cole et al. (2008).  We use a mass 
resolution (i.e. the lowest-mass halo which we trace in our trees) of 
$10^6h^{-1}M_\odot$, which is sufficient to achieve resolved galaxy properties 
for all of the calculations considered in this work.

Within this evolving hierarchy of dark matter, the evolution of the host galaxy
is followed using the {\sc Galform} semi-analytic model of galaxy formation 
(Cole et al. 2000). In particular, we use the implementation of 
Bower et al. (2006) which includes feedback from active 
galactic nuclei and which produces a good match to the evolving stellar mass 
function of galaxies. For the purposes of this work the key results extracted 
from the model are the evolving properties (mass and size) of the disc forming 
at the centre of the dark matter halo. These are computed in {\sc Galform} by 
determining, at each time step, the rate at which hot diffuse gas in the halo 
is able to cool and sink towards the halo centre, where it becomes 
rotationally supported, forming a disc. Simple prescriptions for star 
formation and feedback from supernovae then allow the gaseous and stellar mass 
of the galaxy to be determined. Knowledge of the angular momentum of the disc, 
assumed to have been conserved throughout the process of cooling, collapse and 
disc formation, allows for an estimate of the size to be found, following the 
methods of Mo, Mao \& White (1998).

We run ten different realizations of a MW-like galaxy with a mass of $10^{12} M_\odot$ at $z=0$. Although all models end up with the same mass and virial concentration, the assembly of these haloes is remarkably distinct, a direct manifestation of {\it cosmic variance}. A useful distinction between different halo models can be made by deriving the time at which a significant mass fraction has been accreted. For example, Fig.~\ref{fig:hostevol} includes examples of {\it late-forming} galaxies, which do not accrete 10\% of their present mass until redshit $z\simeq 2$ (e.g solid lines), as well as {\it early-forming} galaxies, which have this mass fraction in place by $z\simeq 5$ (e.g. long-dashed and dotted-dashed lines). 

The disc mass is set by choosing a value for the present disc-to-halo mass ratio $\xi\equiv\big(M_d/M_{\rm vir}\big)_{z=0}$.  Since the baryonic content of galaxies cannot exceed the universal baryon fraction, $\Omega_b/\Omega_m\simeq 0.17$ (Spergel et al. 2007), we consider disc masses within the range $0\le \xi\le 0.2$.
At any given redshift, the disc mass is re-scaled as $M'_d(z)=M_d(z)/ M_d(z=0) \xi M_{\rm vir}$. 
In general the disc component forms at  $z\simgreat 2$ (e.g. {\it solid lines}), although in some extreme cases the formation process can start as early as $z\simeq 5$, (e.g. {\it long-dashed lines}). The disc mass suffers strong fluctuations with time, as shown in the bottom-left panel of Fig.~\ref{fig:hostevol}. 
The formation time of discs plays a key role in the mass evolution of satellites, as discussed in the following Sections.

\begin{figure}
  \includegraphics[width=84mm]{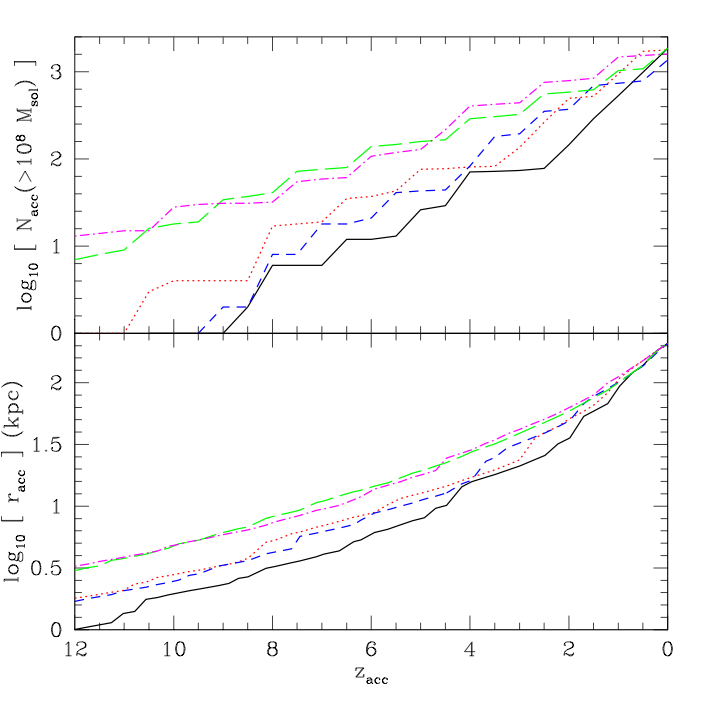}
\caption{ Cumulative number of satellite galaxies accreted with $m_s(z_{\rm acc})\geq 10^8 M_\odot$ as a function of redshift ({\it upper panel}) extracted from the five merger-tree realizations shown in Fig.~\ref{fig:hostevol}. Distance from the host galaxy centre at the time of accretion, $r_{\rm acc}$, as a function of redshift ({\it lower panel}). Note that if orbital decay did not take place, $r_{\rm acc}\sim r_{\rm apo}$ throughout the orbital evolution of satellite galaxies. }
\label{fig:num}
\end{figure}

\subsubsection{The satellite population}

Our merger tree realizations follow the evolution only of subhaloes with masses above $10^8M_\odot$ when they first cross the host virial radius. This choice is motivated by the fact that primordial gas is unable to cool by Lyman-$\alpha$ emission and initiate star formation in haloes with a temperature below $10^4$ K, which approximately corresponds to the above mass threshold. It is typically assumed that lighter subhaloes remain {\it dark} (e.g. Haiman et al. 2000), and thus undetectable through photometric surveys. Unfortunately, determining the number of stars that a given subhalo can form from primordial gas is still a topic of hot discussion (see Tassis et al. 2008; Ricotti 2009 and references therein).

 Our calculations do not include a baryonic component in the accreted subhaloes and may therefore be regarded as {\it pure dark matter models}. Despite the uncertainties in the properties of the stellar components embedded in our satellite models, the following Sections will show that a number of general inferences may be made in reference to the structure of satellite galaxies.

The cumulative number of satellite galaxies accreted at a given redshift, $N_{\rm acc}$, is shown in the upper panel Fig.~\ref{fig:num}. Since the host halo grows by accreting dark matter subhaloes, the number of accreted satellites and the virial mass are closely correlated at any given epoch. An interesting point of note is that a host galaxy with $M_{\rm vir}(z=0)=10^{12}$ is expected to accrete around $\sim 10^3$ satellites that may potentially host a stellar component. This is in remarkable contrast with the number of known satellite galaxies in the MW, which to date adds up to a mere couple of dozens (e.g. Belokurov et al. 2007; Koposov et al. 2009). The apparent mismatch between theoretical predictions and observations is typically referred as the ``missing satellite problem'' (Moore et al. 1999a; Klypin et al. 1999)\footnote{Recent contributions show that the discrepancy between the predicted and observed number of satellites in the Milky Way alleviates after taking into account the detection biases as well as volume corrections of SDSS data (e.g. Tollerud et al. 2008; Koposov et al. 2008).}.

When a subhalo first crosses the host virial radius, it becomes a {\it satellite galaxy}. The inner satellite structure is fixed by applying the virial mass-concentration relation found by NFW. Its orbital parameters are drawn from the distribution of Benson (2005), which was measured from N-body simulations. This distribution gives the radial and tangential velocity components of the orbit. 
By definition, the distance from the host galaxy centre at accretion time, $r_{\rm acc}$, equals the host virial radius, and closely correlates with orbital apocentre (Ludlow et al. 2009). 
Thus, a natural result of the inside-out growth of galaxies is a strong correlation between the accretion time of satellites and their location within the host. 

Indeed, satellites accreted at early times can only be found in the inner-most regions of their parent galaxies. This is illustrated in the lower panel Fig.~\ref{fig:num}, where we show the infall distance as a function of redshift for the merger-tree realizations shown in Fig.~\ref{fig:hostevol}. Two distinct epochs can be distinguished on the basis of the effects introduced by cosmic variance: a late epoch, i.e. during the last 8 Gyr ($z\simless 1$), when the orbits of accreting satellites have a similar distribution of orbital parameters, and an early epoch ($z\simgreat 1$), when the accretion history of the host galaxy is dominated by cosmic variance. This result suggests that the properties of the surviving satellite galaxy population at $z=0$ will show some scatter between different merger trees, and that that scatter will be more evident for satellites that were accreted at early, $z\simgreat 1$, epochs.

\begin{figure*}
  \includegraphics[width=148mm]{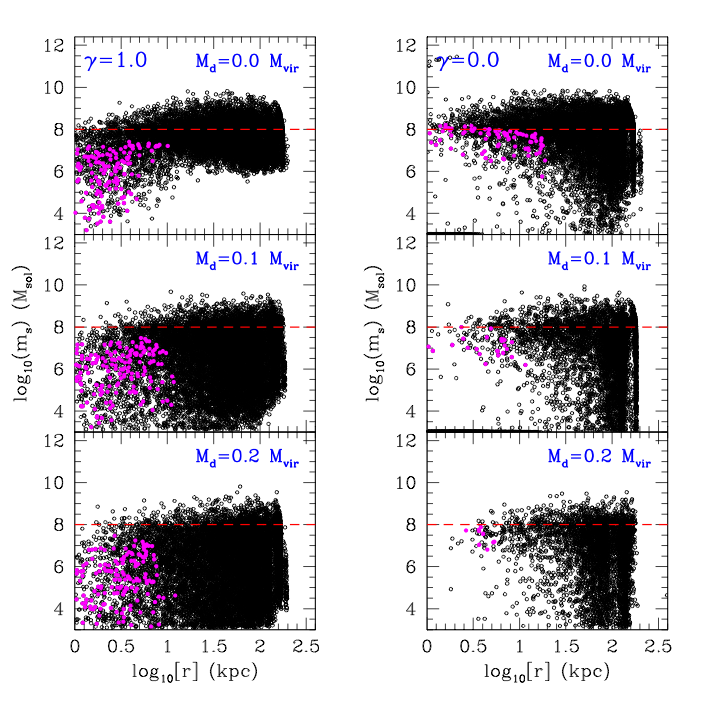}
\caption{ Satellite mass versus Galactocentric radius at $z=0$ for satellite models with $m_s(z_{\rm acc})>10^8 M_\odot$ and different dark matter profiles. Here we stack all the models from {\it ten} different realizations of a host galaxy with $M_{\rm vir}=10^{12} M_\odot$ and $r_{\rm vir}=210$ kpc at $z=0$. Note that all subhaloes had masses above the dashed lines at their time of accretion. Closed symbols denote satellites that were accreted into the host halo with virial masses above $10^8 M_\odot$ at $z>6$ (the so-called {\it fossils of re-ionization}). Note that injecting a baryonic disc in our host galaxy models has the effect of depleting the number of fossils of re-ionization if haloes are cored ($\gamma=0.0$), whereas if haloes are cuspy ($\gamma=1.0$) the main effect is an increase in the mass scatter at a fixed galactocentric radius. }
\label{fig:radmass}
\end{figure*}

\subsection{Spatial distribution of subhaloes}
The destructive action of tides is illustrated in Fig.~\ref{fig:radmass}, where we show the present mass and galactocentric distance of cuspy ({\it left panels}) and cored ({\it right panels}) satellite models extracted from the {\it ten} host galaxy realizations outlined in \S\ref{sec:mtree}. For ease of reference, we mark with dashed lines the minimum mass that satellites must carry as they are accreted into the host in order to be included in our merger trees. Thus in this Figure all satellites with $m_s<10^8 M_\odot$ have lost some fraction of their original mass to tides.

 Fig.~\ref{fig:radmass} illustrates a number of interesting points. The most obvious is that satellites accreted at early times tend to lose a large fraction of their original mass to tides, which is highlighted in this plot by marking the masses and positions of galaxies that fell into the host galaxy at $z>6$ (solid dots). This redshift roughly corresponds to the epoch at which the Universe was completely re-ionized (e.g. Fan et al. 2002), and thus we refer to these galaxies as {\it re-ionization fossils}\footnote{Note that in Ricotti (2009) this term has a different meaning: in particular, it refers to dwarf galaxies with peak velocities below 20 km/s throughout their whole orbital evolution}. It is typically believed that such systems went through a short phase of star formation before they crossed the virial radius of the parent galaxy and lost their primordial gaseous component through ram pressure stripping, a process captured by the N-body simulations of Mayer et al. (2007). The stellar compositions of these systems are thus expected to be distinct from those of the satellite galaxies that continued forming stars after re-ionization and were accreted at a later time. Among the known MW satellites Belokurov et al. (2009) identifies four possible candidates for being fossils of re-ionization: Segue I and II, Bootes II and Coma. Future spectroscopic surveys of MW dSphs will possibly extend this census.

Because the initial orbital apocentres of our satellite models roughly correspond to the virial radius at accretion time, which scales as $R_{\rm vir}\propto (1+z)^{-1}$,  re-ionization fossils are expected to populate the inner-most regions of the host, $r\simless R_{\rm vir}(z=0)/7\simless 30$ kpc (see also Gao et al. 2009). Whether or not re-ionization fossils survive tidal disruption to the present day strongly depends on the inner profile of dark matter haloes, as well as on the mass of the host disc. 

 Fig.~\ref{fig:radmass} shows that the disc component strongly depletes the inner regions of the host galaxy from satellites. Due to their small apocentres, re-ionization fossils count among the satellites most prone to undergo full tidal disruption in $\gamma=0.0$ models. As a result, the number of surviving re-ionization fossils is close to negligible if the inner profile of dark matter haloes is cored. Although cuspy satellite models also shed large fractions of their mass to tides, a bound remnant always survives the shredding process (see Fig.~\ref{fig:mass}). In these models the main effect of the disc component is to lower the present mass (and luminosity) of these systems.

A second point of interest is that the satellite galaxies that have have been acted on by tides extend all the way out to the host virial radius. This is because in CDM most satellites move on highly eccentric orbits (e.g. Benson 2005). Thus, a significant number of galaxies with relatively large apocentres are likely to reach the inner-most regions of the host and experience severe stripping during a Hubble time, as shown in Fig.~\ref{fig:radmass}.

\begin{figure}
  \includegraphics[width=84mm]{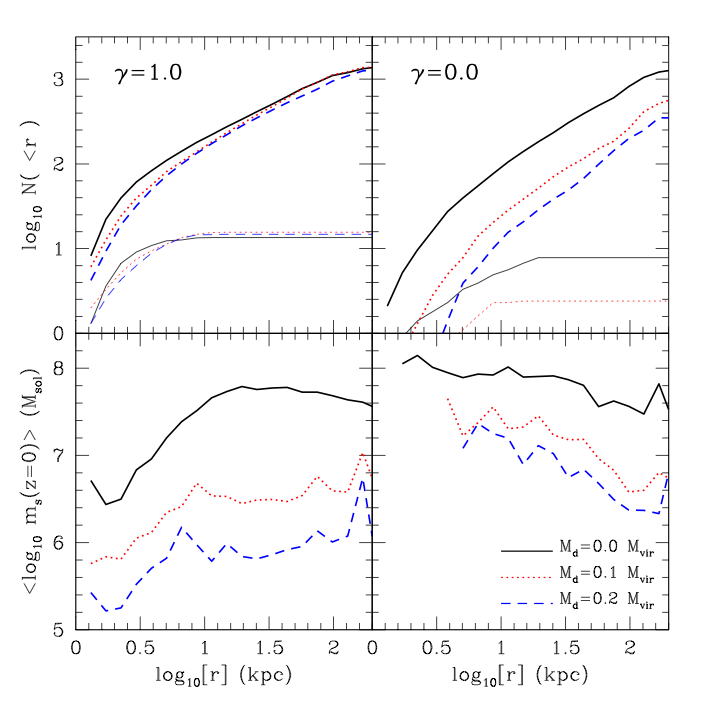}
\caption{ Mean cumulative number of satellite galaxies (upper panels) and mean bound mass (lower panels) as a function of distance from the host centre obtained from the models shown in Fig.~\ref{fig:radmass}. We measure these quantities for two satellite samples: those that were accreted at $z>6$, to which we refer as {\it fossils of re-ionization} (thin lines), and the whole sample, which includes all satellites with masses above $10^8 M_\odot$ at accretion (thick lines). Solid, dotted and dashed lines denote host galaxies with $M_d/M_{\rm vir}=0.0, 0.1$ and 0.2, respectively. Note the present mass and spatial distribution of satellite galaxies strongly depends on the slope of the inner halo profile.}
\label{fig:mrprof}
\end{figure}

In Fig.~\ref{fig:mrprof} we show the mean cumulative number of satellites per realization ({\it upper panels}) and bound mass ({\it lower panels}) as a function of galactocentric distance, as derived from the above models. Thin and thick lines distinguish between fossils of re-ionization and the whole satellite sample, respectively. 
As anticipated above, the cumulative number of cuspy satellites barely depends on disc mass, again illustrating the fact that cusps cannot be completely destroyed by tides. Cored satellites, in contrast, can be fully disrupted by tides, which translates into a progressive decrease of the total number of satellites as more massive discs are assumed. This is particularly obvious for the satellites accreted prior to re-ionization (i.e. $z_{\rm acc}\simgreat 6$). Adopting a fiducial disc mass $M_d=0.1 M_{\rm vir}$ and a cored halo profile, we expect of the order 2--3 fossils of re-ionization in a MW-like galaxy. This number nearly quadruples in halo models with $\gamma=1.0$ and is barely dependent on disc mass. 

A result that may have important observational consequences for the detection of dark matter annihilation signals in gamma-ray surveys refers to the large abundance of satellite galaxies at small galactocentric radii. Cuspy models predict the presence of the order of $100$ satellites within 10 kpc from the host centre with masses that range between $10^3$--$10^8 M_\odot$, a number that barely depends on the disc mass. This number drops by a factor $\sim 10$ in models that assume a cored dark matter profile and a fiducial disc mass of $M_d=0.1 M_{\rm vir}$. Since the gamma-ray flux from a particular source is inversely proportional to the distance squared, these galaxies may provide interesting targets for dark matter annihilation surveys, an issue that we will explore in a separate contribution.

The lower panels of Fig.~\ref{fig:mrprof} show that the variation of the mean satellite {\it bound} mass as a function of galactocentric distance could potentially constrain the inner profile of satellite galaxies. In particular, our models show that the satellite mass is expected to drop at distances $r\simless R_s\simeq 25$ kpc if haloes have a cuspy profile. This is in a marked contrast with cored models, where the effect is the opposite. Here the negative mass gradient arises simply from the depletion of satellites with small orbital apocentres. Satellites that at $z=0$ are located at distances comparable to the disc size happen to be observed by chance near pericentre.

 An interesting point concerns the extreme stripping suffered by a large number of satellites in these models. 
 Pe\~narrubia et al. (2008a,b) estimate that satellite galaxies must lose $\simgreat$90\% of the dark matter envelope before stars begin to be shed to tides. In our models this implies that all satellites in Fig.~\ref{fig:radmass} with masses $\simless 10^7 M_\odot$ have lost a fraction of their stellar components to tides and have associated stellar streams. The fact that tidal stripping is very sensitive to the inner profile of dark matter haloes suggests that surveys of stellar streams in spiral galaxies may put strong constraints on the inner profile of dark matter haloes. Unfortunately, the absence of a stellar component in our satellite models prevents us from exploring this issue in depth.

\subsection{The masses of Milky Way and M31 dSphs}
The results shown in previous Sections suggest that by simply counting the number of satellites around spiral galaxies one could make strong inferences on the inner structure of CDM haloes. However, this requires theoretical models to be able to predict properties of the {\it stellar populations} hosted by dark matter subhaloes and to make comparisons that involve observable quantities, such as galaxy luminosities and surface brightnesses. In practice, this is difficult because, in spite of strong theoretical efforts (e.g. Macci\'o et al. 2009; Cooper et al. 2009), the processes that shape the baryonic properties of satellite galaxies during their formation and later accretion are still rather uncertain.

 A more promising way forward might arguably be to compare the masses of satellites derived from the kinematics of the stellar tracers against theoretical predictions. In MW-size galaxies, the vast majority of satellite galaxies correspond to dwarf spheroidals (dSphs). Although dSph mass estimates are often dominated by large uncertainties (e.g. Kravtsov 2010 and references therein), recent contributions have shown that the satellite mass enclosed within both the projected and tri-dimensional half-light radii of these systems is well constrained and robust to a wide range of halo models and velocity anisotropies (Walker et al. 2009, Wolf et al. 2009). These data exist for most of the known MW satellites and a handful of M31 ones, which altogether encompass half-light radii that span nearly two orders of magnitude in size. 

\begin{figure}
  \includegraphics[width=84mm]{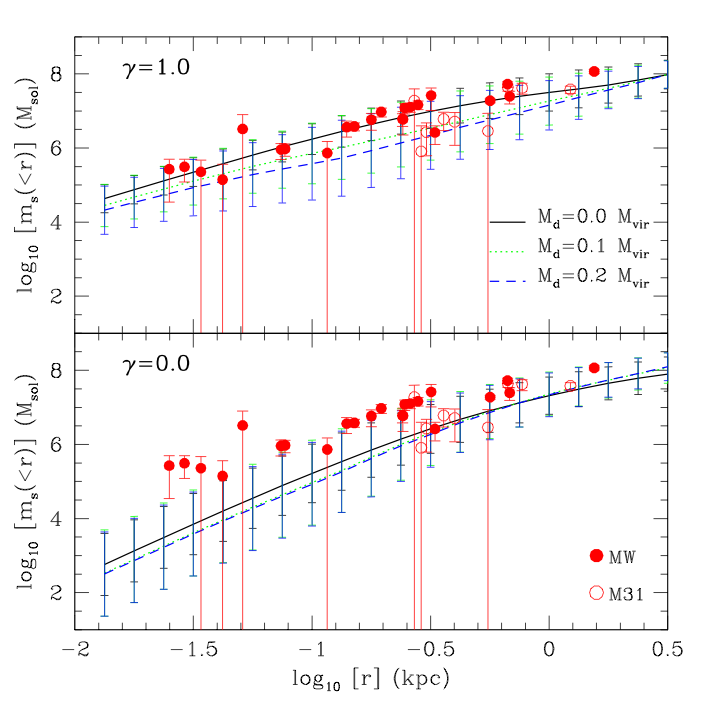}
\caption{ Mass enclosed within fixed radii in the satellite frame for the models shown in Fig.~\ref{fig:radmass}. Lines denote average values and error bars deviations around the mean. Filled and open symbols show, respectively, the masses of the MW and M31 dwarf galaxies with known velocity dispersion estimated at the half-light radius, $R_{\rm half}$, using Walker et al. (2009) formula. To allow a meaningful comparison against observational constraints, only satellite models with  $r_{\rm vir}\geq 2 r$ are considered in the calculation (see text).  }
\label{fig:mrh}
\end{figure}

In Fig.~\ref{fig:mrh} we show the mass estimates for MW (Walker et al. 2009) and M31 (Collins et al. 2009; Kalirai et al. 2009) dwarfs with closed and open symbols, respectively. Lines are derived from the sample of models shown in Fig.~\ref{fig:radmass} by simply measuring the mass of individual satellites enclosed within the range of radii spanned by the observational data, $m_S(<r)$. Error bars attached to the curves denote deviations around the mean, whereas those associated with the symbols represent observational uncertainties. Some of the surviving satellite galaxies in our models have smaller sizes than the largest half-light radii in the sample. To remove this bias, only haloes with $r_{\rm vir}\geq 2 r$ are considered in the calculation of $m_s(<r)$.

This Figure shows two remarkable results. The first one is that cuspy satellite models ($\gamma=1.0$) provide a better description of the masses of Local Group dwarfs than cuspy ($\gamma=0.0$) ones. This is particularly obvious at small radii, i.e. the region populated by ultra-faint dSphs, where the difference between both halo models are largest (see Fig.~\ref{fig:prof}). Although not shown here, we have also explicitely confirmed that this mismatch do not arise from biases between the satellite masses and their galactocentric distance or their time of accretion. 
It is also worth noticing that the deviations around the mean are relatively small taking into accout the large range of bound masses encompassed by our satellite models, and provide a good representation of the scatter shown by the observational data in models where the disc component represents a small mass fraction of the host halo. This result naturally 
accounts for the appearance of a ``universal mass profile'' for dwarf galaxies (Walker et al. 2009). 

The second interesting point concerns the impact of the disc component on the masses of satellite galaxies. Clearly, the mean enclosed mass at a given radius tends to decrease as more massive disc models are adopted. At a given radius, the mean satellite mass can drop as much as a factor $\sim 3$ for the most massive disc components considered here, $M_d=0.2 M_{\rm vir}$. Interestingly, the systematically low masses shown by the handful of M31 satellites with available kinematics may be the telltale evidence of a relatively heavy disc component. In this scenario, the MW and M31 may be embedded in similar-size haloes, but the disc-to-halo mass ratio of M31 would be a factor $\sim 2$ higher than that of the MW. 

Interestingly, by converting K-band absolute magnitudes into stellar masses Hammer et al. (2007) have recently estimated that the disc of M31 may approximately be a factor $\approx 2$ more massive than that of the MW, although these measurement still have large uncertainties. This result, if confirmed, would lend support to the scenario that explains the relatively low mass of M31 satellites as tidal in origin.

Further clues to the mass of M31 disc may be gathered from the faint-end of the dSph population. Our modeling indicates that the mass difference between M31 and MW dSph populations should also be present in dwarfs with half-light radii smaller than those shown here. This CDM-motivated prediction may be verified with present observational capabilities.

\subsection{Mass Threshold for star formation}
In previous Sections a mass threshold at $m_{\rm thres}\equiv m_s(z_{\rm acc})_{\rm min}\geq 10^8 M_\odot$ was imposed to account for the fact that only satellites with masses $m_s\geq m_{\rm thres}$ can form stars (e.g. Heinman et al. 2000). There is, however, some controversy about the value of $m_{\rm thres}$ that defines the minimum galaxy mass (see e.g. Tassis et al. 2008 and Ricotti 2009 for disparate results that place $m_{\rm thres}$ above and below the adopted $10^8 M_\odot$, respectively). 

Because the satellite mass function diverges as $m_s\rightarrow 0$, imposing a mass cut onto the satellite population implies that satellites with $m_s\approx m_{\rm thres}$ dominate in number. As a result, increasing (decreasing) the value of the mass threshold shifts upward (downward) the curves shown in Fig.~\ref{fig:mrh} at all radii. This suggests that the masses of Local Group dwarf galaxies can effectively be used to place constraints on $m_{\rm thres}$. 

\begin{figure}
  \includegraphics[width=84mm]{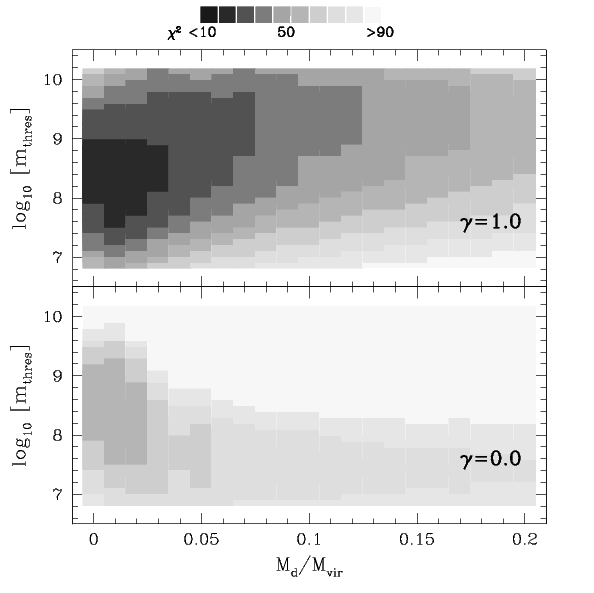}
\caption{ Mass threshold for star formation in satellite galaxies, $m_{\rm thres}$, as a function of host disc mass. Points are grey-coded according to the accuracy of cuspy ({\it upper panel}) and cored ({\it lower panel}) satellite models in describing the mass-size relation shown by MW dwarfs (see Fig.~\ref{fig:mrh}). Note that (i) cuspy models provide a considerably better match to observations and (ii) a mass threshold at $m_{\rm thres}\sim 10^8$--$10^9 M_\odot$ is clearly preferred for a large range of disc masses.}
\label{fig:mthr}
\end{figure}

Fig.~\ref{fig:mthr} shows a crude attempt to estimate the minimum satellite mass above which star formation is expected to occur. To estimate the accuracy of our models in describing the masses of the MW dwarf galaxies shown in Fig.~\ref{fig:mrh}, we apply a $\chi^2$ fit, where
\begin{equation}
\chi^2 \equiv \sum_{i=1,N_{\rm obs}}\frac{1}{N_{\rm obs}}\sum_{j=,N_{\rm sat}} \frac{1}{N_{\rm sat}}\frac{(y_{\rm{obs},i}- y_{\rm{sat},j})^2}{\sigma_{\rm{obs},i}^2};
\label{eq:chi}
\end{equation}
 and $y\equiv \log_{10} M$. The observational variance is $\sigma_{\rm obs}=\sigma_M/(M \log_{10})$, where $M$ denotes dwarf masses estimated at the half-light radius with uncertainties $\sigma_M$. The number of MW dwarfs in Walker et al. (2009) catalogue is $N_{\rm obs}=23$, where $N_{\rm sat}$ is the number of satellite galaxies {\it per realization} with masses above $m_{\rm thres}$. This number strongly depends on the mass threshold, and for a MW-like halo it ranges between $\sim 10^4$ and $10^2$ for $m_{\rm thres}=10^7$ and $10^9 M_\odot$, respectively.

Interestingly, the masses of MW dSphs strongly constrain the mass threshold for star formation in dark matter subhaloes. We find that only those subhalos that were accreted with masses above $10^8$ and $10^9 M_\odot$ have average masses compatible with those in the sample of MW dwarfs with suitable kinematic data. Subhaloes that obey this condition may be identified with ``luminous'' satellites while the rest remain dark, in agreement with a broad number of theoretical expectations (Bullock et al. 2009; Kravtsov 2010 and references therein). 

Comparing both panels we find that, independently of the host disc mass $M_d$, satellite models with a cuspy inner profile provide a better match the MW dwarf galaxy masses than cored models, as expected from Fig.~\ref{fig:mrh}. 

However, there are a number of issues in this comparision that call for caution. For example, the census of MW dwarfs is still incomplete and limited by the surface brightness that observational surveys can reach. It is thus unclear the mass-size relation shown in Fig.~\ref{fig:mrh} holds for the entire satellite population. Also, we have implicitely assumed a lack of correlation between the mass and the stellar properties of dwarf galaxies. Although this is motivated by observational data (e.g. Mateo 1999; Walker et al. 2007; Pe\~narrubia et al. 2008a), self-consistent cosmological models that follow star formation/feedback in subhaloes are required to carry out a more quantitative analysis, as well as to explore in depth the contraints on galaxy formation models provided by the dSph population in the Local Group.

\section{Summary}\label{sec:summary}
We have used N-body simulations to explore the differences that dark matter cusps and cores introduce in tidal mass stripping of satellite galaxies. Our models assume cosmologically-motivated initial conditions, where satellites are dark-matter dominated systems moving on eccentric orbits within a much more massive host galaxy, which is composed of a ``baryonic'' disc embedded in an extended dark matter halo, whose density profile is assumed to be the same as that of satellite galaxies.
Our findings can be summarized as follows

\begin{itemize}

\item Our models show that the resilience of satellite galaxies to tides increases extraordinarily with the slope of the inner halo profile, $\gamma$. In particular, satellite models with $\gamma\geq 1.0$ cannot be fully disrupted by the tidal field of the parent galaxy, and always retain a bound remnant, even after losing more that 99.99\% of mass to tides. In contrast, satellites with cored haloes $\gamma=0.0$ may undergo full tidal disruption if their orbits bring them in the vicinity of the disc.

\item Tides do not affect the inner structure of dark matter haloes, so that $\gamma$ stays constant during the evolution of satellite galaxies. Remarkably, the structural parameters of the satellite halo, e.g. the peak velocity $v_{\rm max}$ and its location $r_{\rm max}$, evolve in a manner that varies depending on the value of $\gamma$, but that is solely controlled by the total amount of mass lost to tides.

\end{itemize}

To examine the differences that dark matter cores and cusps may introduce in the present satellite population surrounding spiral galaxies, we carry out ten different realizations of merger trees that describe the hierarchical formation of a spiral galaxy with properties similar to those of the MW. These realizations are re-run adopting different values of $\gamma$ as well as disc-to-halo mass ratios, $M_d/M_{\rm vir}$. We account for the fact that only a small fraction of subhaloes may be ``luminous'' by only considering those that fall into the host with masses above $10^8M_\odot$. Our main conclusions are the following

\begin{itemize}

\item The net effect of tides is to decrease the average mass of satellite galaxies at all Galactocentric radii. Increasing the mass of the host disc clearly magnifies this effect: the average bound mass drops by a factor $\sim 10$ after injecting a disc component in the host with $M_d/M_{\rm vir}=0.1$, and a factor $\sim 30$ for $M_d/M_{\rm vir}=0.2$ (see also D'Onghia et al. 2009).

\item The present number of satellite galaxies that are accreted before re-ionization puts strong constraints on the inner profile of dark matter haloes. Our models predict that a MW-like galaxy may contain $\sim 15$ of these galaxies orbiting within the central 20 kpc if $\gamma=1.0$, in very good agreement with recent results from the Aquarius team (Gao et al. 2009). The number of these systems, however, quickly drops to zero in models that assume a cored halo profile and host disc component with $M_d \simgreat 0.1 M_{\rm vir}$. These systems may be singled out by anomalous metallicity patterns, and thus represent an interesting observational case to unravel the inner structure of haloes.

\item In models that adopt a cored dark matter halo profile, discs efficiently deplete the inner regions of the host from satellites. In contrast, cuspy satellites can survive in the inner-most regions of spiral galaxies even after losing large fractions of their original mass to tides, which may have important observational consequences for the detection of dark matter annihilation signals in gamma-ray surveys.

\item The relationship between half-light radius and enclosed dynamic mass shown by most Local Group dwarf galaxies, which spans approximately two orders of magnitude in size and three in mass, is successfully reproduced by our models if adopting a $\gamma=1.0$ halo model. In contrast, cored halo models systematically underestimate the masses of the Ultra-Faint dSphs recently discovered in the MW.

\item In order to explain the size-mass relationship exhibited by the MW dSph population, a subhalo mass threshold for star formation is needed. We find that only those subhalos that were accreted with masses {\it above} $10^8$--$10^9 M_\odot$ may have hosted luminous satellites that agree with the observational constraints, while the rest remain dark or beyond the present detection limits. 

\item Our models show that high disc-to-halo mass ratios tend to lower the overall masses of satellite galaxies. A relatively massive M31 disc may explain why many of its dSphs with suitable kinematic data fall below the size-mass relationship established from MW dwarfs. Recent estimates suggest that the disc of M31 may be a factor $\approx 2$ more massive than that of the MW (Hammer et al. 2007), lending support to this scenario.
This explanation is likely to be validated (or challenged) by the results of many ongoing kinematic surveys of M31 dSphs, which will soon determine whether the difference in the size-mass relationship is systematic to the majority of the dSph population surrounding M31.
\end{itemize}

\vskip1cm
JP thanks Stelios Kazantzidis for kindly providing the code used to generate the spherical equilibrium N-body models. AJB acknowledges the support of the Gordon and Betty 
Moore Foundation. We thank the anonymous referee for his/her insightful comments.

{}

\appendix

\section{The semi-analytic code}\label{sec:sa_code}
To follow the dynamical evolution of dark matter subhaloes, we have constructed a simple semi-analytic code that computes their orbit, mass and structural evolution since they cross for the first time the virial radius of the host galaxy to the present. This code permits us to explore in a flexible manner aspects of galaxy formation that, if simulated with the aid of large N-body algorithms, would be notably {\it expensive} in terms of CPU requirements, such as the importance of cosmic variance in making observational predictions, or the effects of a varying the baryonic disc mass on the satellite galaxy population.

\subsection{Computation of orbits}
To calculate their orbits, we treat dark matter subhaloes as point masses that move in a host halo potential whose parameters evolve as shown in \S~\ref{sec:mtree}. At each time step, our code solves the following equation of motion

\begin{equation}
{\ddot {\bf r}}={\bf f}_h +{\bf f}_d + {\bf f}_{\rm df};
\label{eq:eqmot}
\end{equation}
where the first term on the right side of the equation is the host halo force 
\begin{equation}
{\bf f}_h=-G M(<r)/r^2;
\label{eq:force}
\end{equation}
and 
\begin{equation}
M(<r)=4\pi\int_0^r \rho(r')r'^2dr';
\label{eq:mass}
\end{equation}
where $\rho(r)$ has been defined in eq.~[\ref{eq:rho}] and ~[\ref{eq:rhotrun}]. The second term on the right side of the equation refers to the force induced by the baryonic disc, which we approximate as a Miyamoto-Nagai (1975) model (see \S~\ref{sec:mtree}). 

The third term on the right side of the equation is the dynamical friction force exerted by the diffuse sea of dark matter particles on massive satellites orbiting around their host. Several studies of satellite orbital decay have shown that, in spherical systems, Chandrasekhar's formula for dynamical friction (Chandrasekhar 1943) is sufficiently accurate if the Coulomb logarithm is treated as a free parameter to fit to N-body orbits (e.g. Colpi, Mayer \& Governato 1999; van den Bosch et al. 1999). Semi-analytic methods that include Chandrasekhar's dynamical friction have been demonstrated to reproduce accurately the overall dynamical evolution of satellite galaxies in a cosmological context (e.g. Vel\'azquez \& White 1999; Taylor \& Babul 2001; Zentner \& Bullock 2003; Benson et al. 2004; Pe\~narrubia \& Benson 2005) and therefore represent a useful tool for conducting extensive studies of a large parameter space.

 Dividing the background potential into a disc and dark matter halo components, Chandrasekhar's formula can be written as
\begin{eqnarray}
 {\bf f}_{\rm df}={\bf f}_{\rm df, disc}+ {\bf f}_{\rm df, halo}= \\ \nonumber 
 -4\pi G m_s \sum_{i=h,d}\rho_i(r) F(<v_{\rm rel})\ln \Lambda_i \frac{{\bf v}_{\rm rel}}{v_{\rm rel}^3}.
\label{eq:df}
\end{eqnarray}
where $\ln \Lambda_d$ and $\ln \Lambda_h$ are the Coulomb logarithms of the disc and halo components, respectively.
The relative velocity of a satellite with respect to disc and dark matter halo particles is simply ${\bf v}_{\rm rel,d}={\bf v}-{\bf v}_{d,\phi}$, where $v^2_{d,\phi}=R|f_d(Z=0)|$ is the disc circular velocity on the galaxy plane ($Z=0$); and ${\bf v}_{\rm rel,h}={\bf v}$,  where ${\bf v}$ is the satellite velocity vector. 

 For simplicity we assume here that the velocity distributions $F(v)$ of both disc and host halo are Maxwellian and isotropic
\begin{equation}
F(<v_{\rm rel,i})={\rm erf}(X_i)-\frac{2X_i}{\sqrt\pi}\exp[-X_i^2];
\label{eq:fv}
\end{equation}
where $X_i=|v_{\rm rel,i}|/\sqrt{2}\sigma_i$ and $\sigma_i$ is the one-dimensional velocity dispersion, which is defined as $\sigma_i(r)\equiv 1/\rho_i(r)\int_\infty^r \rho_i(r')[f_h(r')+f_d(r')]dr'$.

 Kazantzidis et al. (2004) have examined the inadequacies of the local Maxwellian approximation when applied to NFW halos and, although it may introduce problems to generate halos in perfect equilibrium, the deviations from Gaussianity of the exact distribution function have a negligible impact on the friction term (see their Fig. 4).

 Pe\~narrubia et al. (2004) find that the best fit to self-consistent N-body orbits is obtained for a Coulomb logarithm $\ln\Lambda_h=2.1$, in good agreement with the recent results of Arena \& Bertin (2007). Following Taylor \& Babul (2001), we use $\ln\Lambda_d=0.5$. 
We must emphasize, however, that these quantities are fairly sensitive to numerical aspects, such as particle number and spatial resolution (e.g. Prugniel \& Combes 1992, Wahde \& Donner 1996) as well as to the host's density profile (Hashimoto et al. 2003, Just \& Pe\~narrubia 2005, Read et al. 2006, Arena \& Bertin 2007). It is therefore necessary to note that the decay times of satellites orbiting in dark matter halos are uncertain to a significant degree, since orbital decay rates depend on the value assumed for $\Lambda_d$ and $\Lambda_h$.

\subsection{Mass loss}
Satellite galaxies may lose a fraction of their original mass to tides if their orbits bring them into the inner regions of the parent galaxy. A simple way to estimate the bound mass fraction of a satellite galaxy is to define the tidal radius as the distance from the satellite centre at which the force due to the satellite's self-gravity and that from the external tidal force balance. For satellites moving on circular orbits, one can estimate that radius as (King 1962)
\begin{equation}
R_t\approx \bigg ( \frac{G m_s}{\omega^2 -\rm d^2 \Phi_h/\rm d r^2} \bigg)^{1/3};
\label{eq:rt}
\end{equation}
where $\omega$ is the angular frequency of the orbit and $\Phi_h$ is the potential of the host halo. In calculating $\rm d^2\Phi_h/\rm d r^2$, we will average over the asphericity of the potential due to the disc component, so that
\begin{equation}
\rm d^2\Phi_h/\rm d r^2= \rm d(-GM(<r)/r^2)/\rm d r.
\label{eq:d2phi}
\end{equation}

If the satellite follows an eccentric orbit, one can still use eq.~[\ref{eq:rt}] to calculate the instantaneous tidal radius at orbital pericentre, where tidal forces are strongest.

\begin{figure}
  \includegraphics[width=84mm]{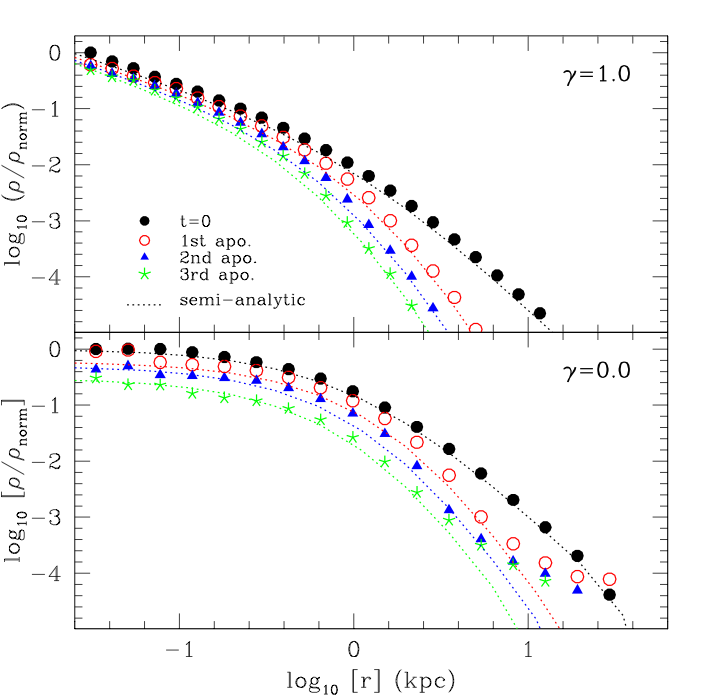}
\caption{ Evolution of the density profile of our satellite galaxy model moving on a highly eccentric orbit ($r_{\rm peri}:r_{\rm apo}=1:100$) at different orbital apocentres (Fig.~\ref{fig:rhoevol}). Profiles are arbitrarily normalized by the quantity $\rho_{\rm norm}$, which is kept fixed in each panel to emphasize the density drop resulting from tidal mass stripping. Dotted lines correspond to our analytical model predictions (see text). }
\label{fig:rhotest}
\end{figure}

\subsection{Evolution of the density profile}
As Fig.~\ref{fig:rhoevol} illustrates, tidal mass stripping lowers the underlying density of dark matter subhaloes at all radii, but more effectively in the outer-most regions. This process has been studied in some detail by Pe\~narrubia et al. (2008b, 2009), who showed that the outer profile of dark matter haloes that lose a fraction (even as small as 10\%) of its mass to tides can be well fitted by a power-law profile that scales as $r^{-5}$ as $r\rightarrow\infty$. In adopting the density profile outlined by eq.~[\ref{eq:rho}], this means that $\beta\rightarrow 5$. It is interesting to note that stripped haloes have $\beta>3$, so that the resulting cumulative mass profile does {\it not} diverge at large radii. Remarkably, in the inner-most regions the dark matter halo profile is barely altered by the action of tides. In practical terms this means that $\gamma$ as well as $\alpha$ can be assumed to stay nearly constant throughout the dynamical evolution of our subhalo models.

Furthermore, the evolution of the structural parameters of a subhalo as it is stripped can be accurately described by {\it tidal tracks}, i.e. mono-parametric functions that only depend on the fraction of mass lost to tides (see Fig.~\ref{fig:vmrm}). 

We can use the above information to describe the changes induced by tidal mass stripping on the density profiles of individual dark matter subhaloes as a function of their bound mass fraction. This is done in a fairly simple way: (i) for $m_s/m_s(t=0)\leq 0.9$ the indices of eq.~[\ref{eq:rho}] are modified as $(\alpha, \beta, \gamma)\rightarrow (\alpha, 5, \gamma)$. (ii) For a given value of $m_s/m_s(t=0)$ we calculate $(r'_{\rm max}, v'_{\rm max})$ using eq.~[\ref{eq:g}]. (iii) The new scale radius $r'_s$ is then calculated from the condition $\rm d V_c^2/\rm d r  |_{r=r'_{\rm max}}=0$, (iv) whereas $\rho'_0$ is obtained using the definition $V_c(r'_{\rm max})\equiv v'_{\rm max}$. 

This simple method permits us to follow the evolution of the density profile of satellite galaxies as they orbit within their host with an appropiate degree of confidence. For example, symbols in Fig.~\ref{fig:rhotest} illustrate the density profiles shown in Fig.~\ref{fig:rhoevol}, whereas dotted lines show the profiles derived from the above recipes. Except for the excess of mass visible at large radii in some cored models, which corresponds to unbound material on its way to escape from the subhalo, our semi-analytic algorithm successfully matches the evolved density profiles, even those of subhaloes that have lost a large fraction of their initial mass. This code will permit us to explore in a very efficient way aspects of galaxy formation that would otherwise require an extensive use of supercomputers, such as accounting for cosmic variance to make theoretical predictions, or the effects that varying the host galaxy disc-to-halo mass ratio would induce on the surviving satellite galaxy population.

\subsection{Tests}
\begin{figure}
  \includegraphics[width=84mm]{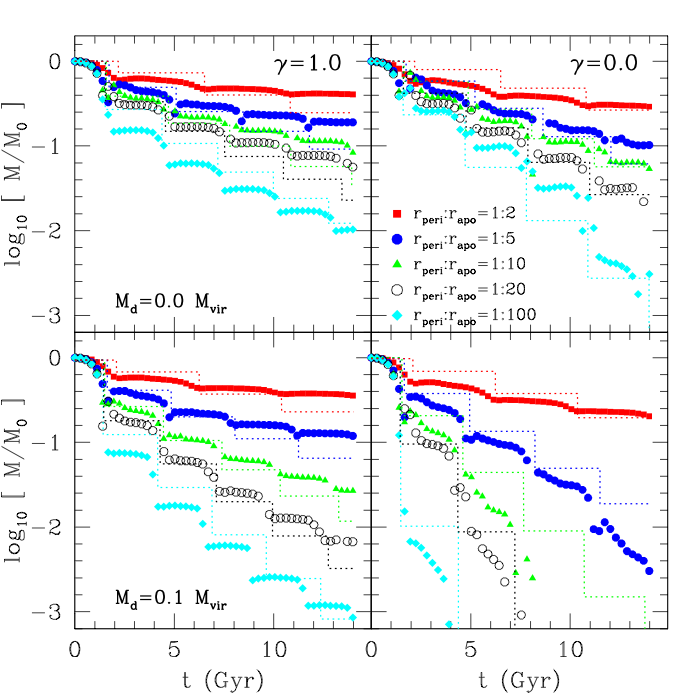}
\caption{ Mass evolution of our satellite galaxy models moving on different orbits. Dotted lines correspond to our analytical model predictions (see text). }
\label{fig:inditest}
\end{figure}

We have used the N-body simulations of \S\ref{sec:tidev} to test our mass loss algorithm, as well as the analytical description of the effects that tides cause on the density profiles of dark matter haloes. The results are shown in Fig.~\ref{fig:inditest}, where we plot the mass our subhalo models as a function of time as derived from our N-body models (see Fig.~\ref{fig:mass}), as well as from the above semi-analytical method (dotted lines). The agreement between both techniques is reasonable, taking into account the extreme mass-loss events considered here and the large range of orbits and structural parameters adopted in this study. More importantly, our simple semi-analytical description is able to capture the strong effect that discs have on the mass evolution of dark matter subhaloes depending on the inner shape of the density profile, which will permit us to explore the effects that baryonic discs have on the satellite galaxy population depending on whether haloes are cuspy or cored.

\end{document}